\documentclass[12pt,twoside]{article}   
\usepackage{epsfig,times}  
\usepackage{color}

\begin{document}
\thispagestyle{empty} 


 \renewcommand{\topfraction}{.99}      
 \renewcommand{\bottomfraction}{.99} 
 \renewcommand{\textfraction}{.0}


\newcommand{\nc}{\newcommand}

\nc{\qI}[1]{\section{{#1}}}
\nc{\qA}[1]{\subsection{{#1}}}
\nc{\qun}[1]{\subsubsection{{#1}}}
\nc{\qa}[1]{\paragraph{{#1}}}

\def\qbu{\hfill \par \hskip 6mm $ \bullet $ \hskip 2mm}
\def\qee#1{\hfill \par \hskip 6mm #1 \hskip 2 mm}

\nc{\qfoot}[1]{\footnote{{#1}}}
\def\qL{\hfill \break}
\def\qpar{\vskip 2mm plus 0.2mm minus 0.2mm}
\def\tvi{\vrule height 12pt depth 5pt width 0pt}
\def\qtvi{\vrule height 2pt depth 5pt width 0pt}
\def\qth{\vrule height 15pt depth 0pt width 0pt}
\def\qtb{\vrule height 0pt depth 5pt width 0pt}

\def\qparr{ \vskip 1.0mm plus 0.2mm minus 0.2mm \hangindent=10mm
\hangafter=1}

\def\qdec#1{\par {\leftskip=2cm {#1} \par}}
%
\def\qbfb#1{{\bf\color{\blue}{#1} }}

\def\qdpt{\partial_t}
\def\qdpx{\partial_x}
\def\qddpt{\partial^{2}_{t^2}}
\def\qddpx{\partial^{2}_{x^2}}
\def\qn#1{\eqno \hbox{(#1)}}
\def\qds{\displaystyle}
\def\qw{\widetilde}
\def\qmax{\mathop{\rm Max}}   
\def\qmin{\mathop{\rm Min}}   

\def\qs#1{{\bf \color{blue} \LARGE {#1}}\quad }

\def\qv{\vskip 0.1mm plus 0.05mm minus 0.05mm}
\def\qhu{\hskip 1mm}
\def\qhv{\hskip 3mm}
\def\qvv{\vskip 0.5mm plus 0.2mm minus 0.2mm}
\def\qhw{\hskip 1.5mm}
\def\qleg#1#2#3{\noindent {\bf \small #1\qhw}{\small #2\qhw}{\it \small #3}\qv }


\centerline{\bf \Large  A joint explanation of infant and old age mortality}
\vskip 5mm

\centerline{Peter Richmond$ ^1 $, Bertrand M. Roehner$ ^2 $}

\vskip 10mm

{\it \large Preprint version of the paper which appeared as:\qL
``Journal of Biological Physics'' 47,131-141 (2021)}

\vskip 10mm

{\bf Abstract} \qL
Infant deaths and old age deaths are
very different.
The former are mostly due to severe congenital
malformations of one or a small number of
specific organs. On the contrary,
old age deaths are largely the outcome of a long process of
deterioration which starts in the 20s and affects
almost all organs. \qL
In terms of age-specific death rates,
there is also a clear distinction:
the infant death rate falls off
with age, whereas the adult and old age death rate increases
exponentially with age in conformity with Gompertz's law. 
An additional difference is that whereas aging and old age 
death have been extensively studied, infant death received
much less attention. To our 
knowledge the two effects have never been inter-connected.
\qL
Clearly, it would be satisfactory to
explain the two phenomena as being two variants within
the same explanatory framework. In other words,
a mechanism
providing a combined explanation for the two forms
of mortality would be welcome. This is the purpose
of the present paper. \qL
We show here that the same biological
effects can account for the two cases provided there
is a difference in their severity: death triggered by
isolated lethal anomalies in one case and widespread 
wear-out anomalies in the second. We show that quite generally
this mechanism leads indeed, respectively,
to a declining and an
upgoing death rate. Moreover, this
theoretical framework
leads to the conjecture that the severity of
the death effects, whether in infancy or
old age, are higher
for organisms which are comprised of a larger
number of organs.
Finally, let us observe that the main focus of the paper
is the drastic difference of the age-specific death rates
(i.e. decreasing versus increasing) because this
difference is found in many species, whereas the question
of the best fit
(e.g. Gompertz versus Weibull) is rather specific to human
mortality.

\vskip 10mm

1: School of Physics, Trinity College Dublin, Ireland. \qL
Email: peter\_richmond@ymail.com
\qpar

2: Corresponding author.\qL
Institute for Theoretical and High Energy Physics (LPTHE),
Pierre and Marie Curie Campus, Sorbonne University, National
Center for Scientific Research (CNRS), Paris, France. \qL
Email: roehner@lpthe.jussieu.fr

\vfill \eject

\qI{Introduction: infant versus old age mortality}

In this paper we consider the shape of the curves of
death rates%
\qfoot{We use the standard definitions of death rates,
namely: $ \mu (x)=\Delta y/(\Delta x\times y) $ where 
$ \Delta y $ is the number of deaths in a given age
interval of size $ \Delta x $ and
$ y $ is the size of the population at
the beginning of the age interval under consideration.
With this definition $ \mu(x) $ is the probability (per
unit of time)
that a person who has reached age $ x $, will die in the 
subsequent age interval, see appendix A.}%
as 
a function of age.

Deaths in infancy versus old age death
can be characterized in two ways [1,2]: 
(i) their age-specific death rates
(see Fig.1) (ii) the biological processes which are 
at work (wear-in versus wear-out as discussed below.)

\qA{Shape of the age-specific death rate} 

In infancy the death rate {\it decreases} with age
whereas in old age it {\it increases} (Fig.1).
%
\begin{figure}[htb]
\centerline{\psfig{width=15cm,figure=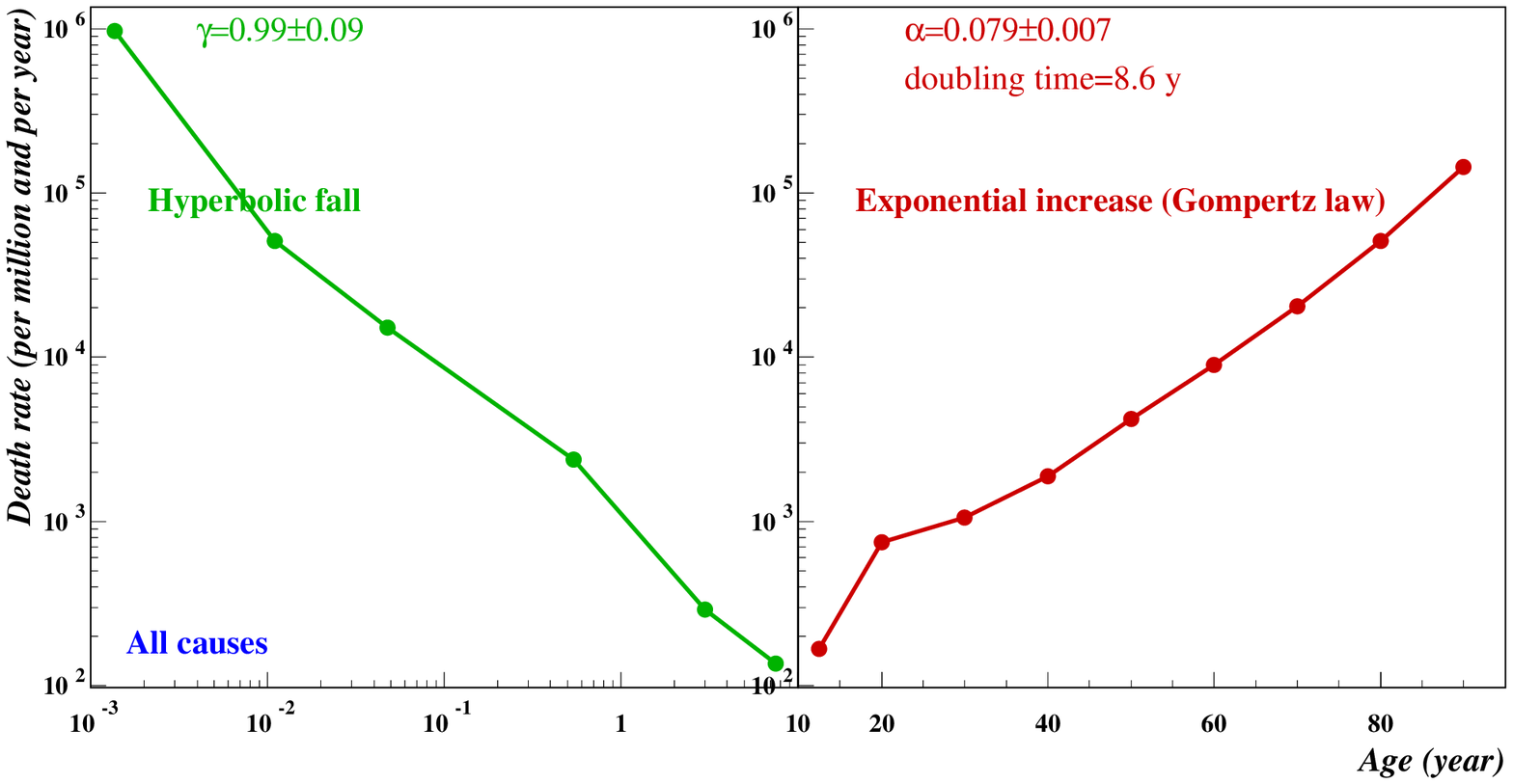}}
\qleg{Fig.1: Infant versus old age human mortality.}
{The data are for the US over the period 1999-2016.
Between birth and the age of 10 (note the log-log scale) 
the infant mortality rate
falls off as a power law: $ \mu_b=A/x^{\gamma} $ where the exponent
$ \gamma $ is usually of the order of 1. After the infant phase
comes the aging phase (note the lin-log scale) during which
the death rate increases exponentially $ \mu(x)=\mu_0\exp(\alpha x) $
in agreement with Gompertz's law.}
{Source: Wonder-CDC data base for detailed mortality.}
\end{figure}

In medical terminology infancy refers to new born under
one year of age. However, as the decrease of the death rate
continues until
the age of 10, it seems appropriate to extend the meaning
of the term to the whole age interval over which the death
rate is decreasing. This is what was done in the two papers
cited above and we use the same terminology here.\qL
For humans, the increase of the death rate is described
by the well-known law of Gompertz [3].
This law can be summarized by saying that
the death rate doubles approximately every 10 years of age. 

\qA{Wear-in versus wear-out} 

In the terminology of reliability studies,
infant mortality is described by a {\it wear-in} process,
that is to say a phase during which the organs 
of the new-born start to work which results in 
the elimination of the organisms which are beset with
an organ which does not work appropriately.\qL
On the contrary, old age death  is described as a
{\it wear-out} process in which all organs experience
damages due to continuous use. The lungs
catch less oxygen, the bones become more fragile,
the arteries become less elastic, and so on. 
Death occurs eventually due to the failure of a crucial 
organ (usually the heart and lungs)
but actually this failure is favored by the degradation
of the whole organism. For instance, when the arteries
become less elastic, when the lungs become less effective,
it becomes more difficult for
the heart to ensure blood circulation. This means that
a heart failure does not come about in isolation but
rather in relation with the wear of other organs.
\qpar

The purpose of this article is to show how the feature
(1) results from the feature (2). However, before coming to 
that we wish to explain how our study fits into
the broad framework of aging and senescence studies.
\qpar

Why do we think that infant mortality is an essential component
in the understanding of aging? There are several reasons that are
developed below.

\qI{The key-role of congenital malformations}

\qA{The real challenges of aging models}

Not surprisingly, the modeling of aging and senescence
has received great attention. A comprehensive review 
can be found in two
papers by Leonid Gavrilov and and Natalia Gavrilova [4,5].
Readers will find in these papers a comprehensive and
very readable account of the literature of aging models%
\qfoot{Whereas these authors share our approach 
based on reliability
science, and whereas infant mortality is a standard notion
in reliability, they devote only one page (in a total
of 58 for the two papers) to the question of infant mortality.
This disproportion reflects the overwhelming predominance
of aging and senescence studies.}%
.
In the following subsections we explain why infant mortality
is a simpler and more fundamental effect than old age mortality.
\qA{Great diversity in the shape of the death rate in old age}

Gompertz (1825) was the first to propose that the rapid rise in death
rate of humans as
they aged prior to death followed and exponential law. Other authors
have 
since advocated a description based on Weibull functions. 
See for example [6,7,8].
An interesting review
of the implications can be found in Roberto Ricklefs and 
Alex Scheuerlein [9]. Of
especial note here is the proposal that the Weibull form holds for death by
particular causes whereas the Gompertz form holds for death by all
causes. 
We shall
return to the point later. However neither form predicts the
hyperbolic 
fall in
deathrate which happens in early years. Neither the Weibull 
nor Gompertz forms
yields insight into data for species which exhibit a levelling off of 
the death rate at very high ages.
\qpar

Across species the hyperbolic decrease of the death rate in the
infancy phase appears to be a phenomenon that is more
widespread than its exponential (i.e. Gompertz-like) increase in
adulthood. Indeed, there is much more
diversity in old age death rates than in infancy death rates;
see in Berrut et al. (2016) the graph based on zoo species. 
\qpar

In
addition for some species documented in [5, p. 18 and 33]
there is a marked effect of leveling-off in old age.
For instance, house flies have a maximum life span of 40 days but around
the age of 15 days
the exponential growth of the death rate is replaced by a section
where it is practically flat. 
\qpar

In humans the dominant
diseases in old age are not the same nowadays as one century ago.
Presently, there is a
prevalence of heart, cancer and Alzheimer's disease whereas around 1900
infectious diseases were still common.
Thus, with organisms being confronted to different challenges,
one should not be surprised 
to see changes in the shape of the death rate in old age.
\qpar

Finally, the individuals who reach old age were ``filtered'' and
selected by the diseases to which they were confronted. If one could
observe the signature of the immune system one would see that 
the immune system of persons of old age is not the same in 2020
than in 1900, and also not the same in developing countries
than in developed countries.

\qA{Common  characteristics  of embryonic and infancy death rates}

It has been shown recently [10] that for zebrafish%
\qfoot{For a study of embryonic death rates, zebrafish have two
great advantages. (i) As for most species of fish, fertilization
of the eggs
occurs outside of the body of the female (ii) The shell of the eggs
is transparent. Taken together, these two features imply
that one can observe the embryos immediately
after fertilization, something that is impossible
either for humans, birds or rotifers.}
the
embryonic death rate is by far highest at the beginning of the
embryogenesis, an observation which suggests that most of these
deaths are due to mistakes in the manufacturing processes
of the oocyte (femelle egg) and sperm cell. If 
instead the deaths would be due to mutations during
the embryogenesis they would be uniformly distributed
or even (through a cumulative effect) concentrated in the
late phase of embryogenesis.
\qpar

For humans embryonic deaths would in medical terminology
be referred to as fetal deaths or still births. In statistical
releases this pre-birth mortality is treated apart from
infant mortality mostly because there is a great uncertainty
about fetal death data. However, from a biological 
perspective, infant mortality is nothing but the continuation
of fetal death, albeit in more severe form due to
disconnection of the link with the organism of
the mother.
\qpar

Similarly, the infant death rate is by far highest 
immediately after birth. By the same argument, it appears
that most of these deaths are due to faults in the
manufacuring of the embryo. For instance, in mamals 
lung malformations are
without consequence as long as the fetus 
receives its blood from the mother but they will lead to death 
as soon as this connection is interrupted.
Embryonic and infant deaths along with the malformations which are not 
immediately lethal give us global information about the underlying
manufacturing processes.

\qA{The effects of congenital defects and of aging occur jointly}

At first sight it may seem that the infant death 
rate can be easily described and explained 
through the process of elimination
of individuals with malformations. Clinical data show
that in the first weeks after birth most of the deaths
are due to congenital anomalies (percentage data are given 
in [1]. When the most serious
malformations have been eliminated the rest of the cohort
is less likely to die.
\qpar

However, the previous explanation is not really
satisfactory for the following reason.
In fact, deaths due to congenital anomalies 
are not limited to young age but
continue during the whole life. For instance, a congenital
defect of heart valves may be of no consequence
until the age of 60 or 70 when the defect becomes more
serious because the valve's leaflets
become stiffer, see [11].
\qpar

In other words, the wear-in and wear-out processes
should not be seen as occurring successively but rather
simultaneously; it is their strength, not their existence,
which changes in the course
of time. Immediately after birth, wear-in is completely
dominant, whereas in old age it is wear-out which
is predominant. I short, taken alone the elimination
of congenital malformations cannot explain the decrease 
of the death rate. In order to make it work we 
need to define both wear-in and wear-out more
precisely. 
\qpar

In the next section we will use the feature
already mentioned above, namely that the infant mortality 
is usually due to a congenital defect in one important
organ (e.g. heart, lung, brain, liver, and so on)
whereas the wear-out is due to parallel degradation
of various important organs.

\qI{Modeling the wear-in and wear-out processes}

\qA{Decomposition into vital organs}

The first step is to decompose any organism
into its vital organs. For instance, Fig.2 shows
a decomposition into 4 organs, that could be
heart, lung, brain and temperature regulation.
\qpar

%
\begin{figure}[htb]
\centerline{\psfig{width=12cm,figure=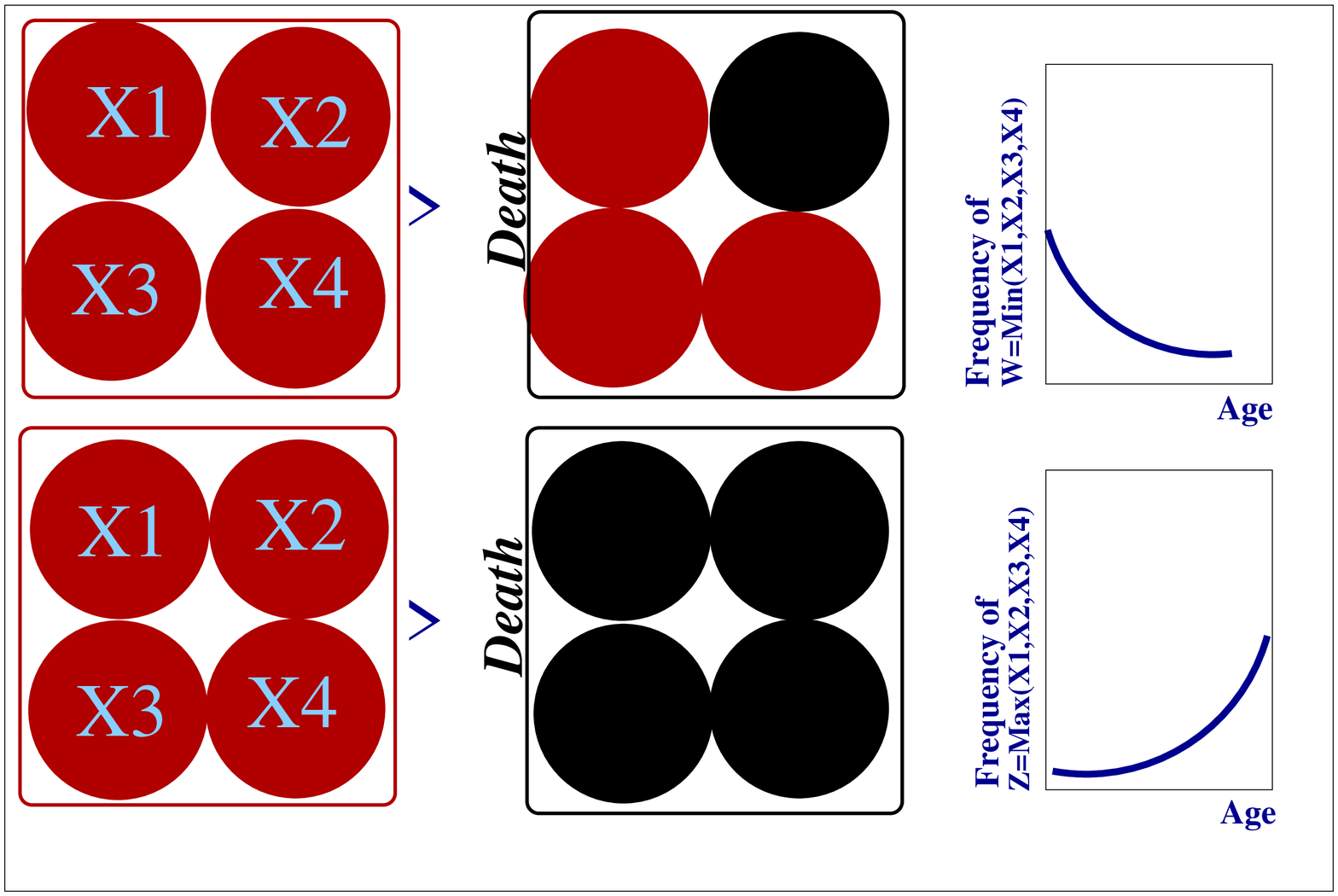}}
\qleg{Fig.2\quad Decomposition of an organism into vital organs
and difference between wear-in and wear-out mechanisms.}
{The first line shows a wear-in death. It is the
consequence of the failure of a single vital organ.
The second line shows a wear-out death as the
consequence of uniform deterioration of all
vital organs. The graphs on the right-hand side
show the implications of these mechanisms in terms
of age-specific death rates: decreasing for wear-in
as observed in infant death,
increasing for wear-out as seen in old-age death.}
{ }
\end{figure}

\qA{Description of the organs' state by random variables}

Secondly, we must find a way to describe
mathematically whether each organ (as well as the whole
organism) is alive. We do this by defining for each part
a random variable $ X_i $ which is its age at failure. 
In the case of humans we could make this description
fairly realistic by giving to the $ X_i $ values
from interval $ (0,125) $ 
for it should be remembered that 125 years is an upper bound
of human life%
\qfoot{This was shown to be a consequence of Gompertz's law
in [12].}%
.
However, as we do not wish to
restrict ourselves to only the human species, we will normalize
the interval of the $ X_i $ to $ (0,1) $ where 1 represents
the maximum life span of the species.
\qpar

Now comes the most important step which is to describe
the wear-in and wear-out mechanisms. 
Let us begin with the
simplest case which is the wear-out mechanism.

\qA{Wear-out}

The fact that the death of an individual
occurs when the last surviving organ fails is expressed
by saying that if $ X_1=0.5,\ X_2=0.3,\ X_3=0.7,\ X_4=0.1 $
(the $ X_i $ denote the age at death of vital organs as
shown in Fig.2),
then the age of death represented by the random variable
$ Z $ will be $ Z=0.7 $, in other words:
 $$ Z=\hbox{Max}(X_1,X_2,X_3,X_4) $$

For the sake of simplicity we assume that the $ X_i $ are
independent and identically distributed random variables.
This assumption has the merit of making the analytical
derivation possible. However, in specific
applications one can take realistic distributions based
on clinical data.
\qbu If $ f(x) $ and $ F(x) $ respectively
represent the density function and the cumulative
distribution function of the $ X_i $, what will be the
density function, $ f_Z(x) $, of $ Z $? 

Let us first consider
the case of only two organs.
$$ F_Z(x) = P\{Z \le x\}=P\{\hbox{Max}(X_1,X_2) \le x\} 
=P\{X_1 \le x \hbox{ and } X_2 \le x\} $$

Now, the fact that $ X_1 $ and $ X_2 $ are independent
means that:
$$ P\{X_1\in A \hbox{ and } X_2\in B\}=P\{X_1\in A\}P\{X_2\in B\} $$

where $ A $ and $ B $ are two subsets of the set of real numbers.
Therefore:
$$ F_Z(x) = P\{X_1 \le x\}P\{X2 \le x\}=\left[ F(x) \right]^2 $$

which, by differentiation leads to the density function of $ Z $:
$$ f_Z(x)=2F'(x)F(x) =2f(x)F(x) $$

For $ p $ parts instead of only two, one gets similarly:
$$ F_Z(x)=\left[ F(x) \right]^p \Rightarrow 
f_Z(x)=pf(x)\left[ F(x) \right]^{p-1} \qn{1} $$

In order to see what is the shape of this function we consider
the simple case of a random variable with a uniform
density over the interval $ (0,1) $; in this case:
$$ \hbox{for } x\in (0,1): \ f(x)=1,\ F(x)=x $$

Thus, 
$$ \hbox{for } x\in (0,1): \ f_Z(x)=px^{p-1} $$ 

This function is shown in Fig.3b for $ p=2,4,8,15 $
We see that it
is a fast {\it increasing} function of age.
According to the analytical expression
$ f_Z(x) $ is a power law function.
This is consistent with a Weibull distribution 
but when $ p $ becomes
large it has the shape of an exponential
(as shown in Fig.2b for $ p=15 $),
a result which is 
qualitatively consistent with Gompertz's law according
to which
the probability of death increases exponentially
with age.
\qpar

At this point it is interesting to recall earlier comments
that the Weibull form appears consistent with deaths from particular
causes. 
Here we find that the more 
elements are present the
closer the death rate approaches a Gompertz exponential form. 
There are ways this
prediction could be checked. One is to assess death rates from 
increasingly complex
synthetic systems; another albeit more difficult could be 
to explore death rates of
increasingly complex biological systms from single cell upwards. 
This may however be
difficult since complexity in biology brings in multiple organisms 
but examining
death rates from all causes should lead ultimately to the Gompertz form.

\qA{Wear-in}

For the example considered above wear-in death would mean
that the age of death is: $ W=0.1 $, i.e.:
$$  W=\hbox{Min}(X_1,X_2,X_3,X_4) $$

For the distribution fonction $ F_W(x) $ we can write:
$$ F_W(x)=P\{W \le x \}=1-P\{W > x \}=
1-P\{ \hbox{Min}(X_1,X_2) > x \}=
1-P\{X_1>x \hbox{ and } X_2>x \} $$

Again using the independence property, one gets:
$$ F_W(x)=1-P\{X_1>x \}P\{X_2>x \}=1-\left[ 1-F(x) \right]^2 $$

Then, as above, this result generalizes to:
$$ F_W(x)=1-\left[1-F(x) \right]^p \Rightarrow f_W(x)=
pf(x)\left[1-F(x) \right]^{p-1}    \qn{2} $$

For the case of uniform random variables, one gets:
$$ \hbox{for } x\in (0,1): \ f_W(x)=p(1-x)^{p-1} $$ 

which means that the probability of death is 
a {\it decreasing} function of age, consistent
with what is expected for infant mortality. The decrease
is ilustrated in Fig.3a for $ p=2,4,8,15 $.

\qA{Special cases}

One can gain an intuitive understanding of the theoretical
framework by considering a number of special cases.
\qpar

Firstly, we can consider
the extreme case of an organism with a high number of components,
say one million or if you prefer a number $ p $ which tends to infinity.
Then, intuitively,
the wear-in assumption gives a probability of death equal to 1
because for such a large number of components there will always be
one which will fails almost immediately.\qL
For the same reason 
the wear-out assumption gives a probability of death equal
to zero for almost all ages because it will take a
very long time to eliminate all and any components.
\qpar

The other extreme case is an organism with only one vital component.
Then, obviously, the two assumptions should give the same result.
Indeed, the formulas (1) and (2) give: $ f_W(x)=f_Z(x)=f(x) $.
Here the shape of $ f_W $ and $ f_Z $ is completely
determined by $ f(x) $  which can have any shape, whether increasing
or decreasing.
\qpar

As the number $ p $ of components increases
the factor $ (1-F)^p $, which is a decreasing
function will become more and more predominant.
Similarly, for old age, the factor $ F^p $, which is
an increasing function, will become predominant
when $ p $ increases.
\qpar

In summary, this discussion makes clear that
equations (1) and (2) do not describe only
one model but, by playing with $ p $ and $ f(x) $,
they can describe a whole spectrum of cases.
It is in this sense that the model is really
predictive.

\qA{Graphs of infancy and old age death rates} 

The density functions of the variables $ W $ and $ Z $ 
give the infant and old age death rates respectively
(Fig.3a,b).

%
\begin{figure}[htb]
\centerline{\psfig{width=16cm,figure=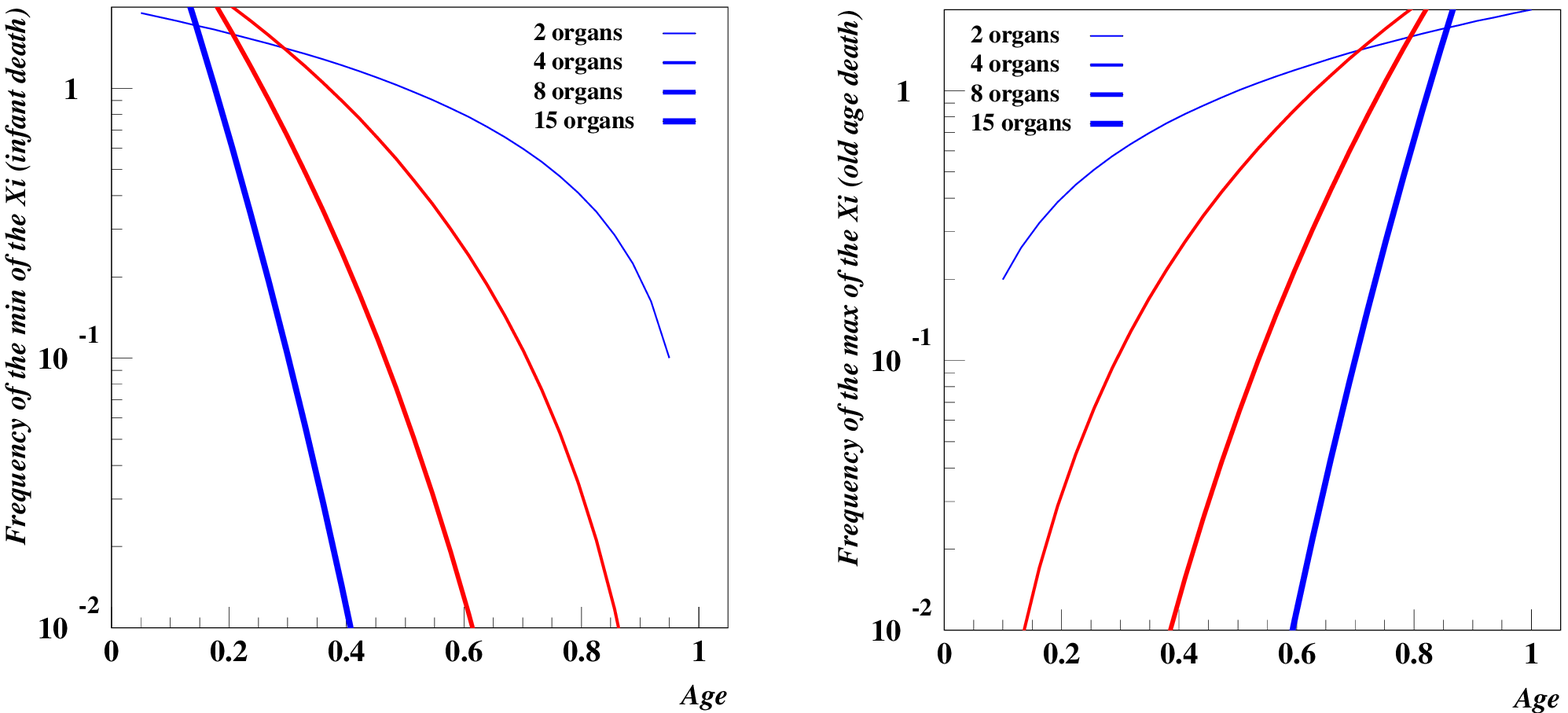}}
\qleg{Fig.3a,b \quad Graph of the density functions of the
Min and Max of a set of random variables.which represent
the age-specific death rate in young and old age respectively.}
{When the number 
of organs becomes large, the death rates
become exponential.}
{ }
\end{figure}

The graphs suggest the conjecture that the 
larger the number of
components, the steeper the death rates with respect to age.
When $ p \rightarrow \infty $  it can be seen directly on
the formulas that $ f_W $ falls vertically from $ pf(0) $ to 
zero, whereas  $ f_Z $ jumps vertically from
zero to $ pf(1) $.
\qpar

As technical systems have often many well-defined, separate
components it should be possible to test our conjecture.
For instance microprocessors comprising a large
number of chips should have steeper death rate curves than
those with only a small number.of chips.
Unfortunately, for such technical systems 
there are almost no life-time data publicly
available, probably for reasons of commercial 
confidentiality.

\qI{Conclusions}

We have proposed two paradigms of death: \qL
(i) single-organ death
which occurs through the failure (e.g. due to congenital
malformation) of one crucial organ, \qL
(ii) multi-organ death which comes about through the
deterioration of almost all organs.
\qpar

At this point a distinction should be made
between the underlying cause of death (e.g. cancer
or an infectious disease) and the immediate cause
of death. It is the former which is of interest
and is reported in the death statistics by cause
of death. On the contrary, the immediate
cause of death is almost always the same,
namely a heart failure. \qL
For instance, a liver cancer will
lead to blood poisening which makes the heart
unable to perform its function properly.
It is because other organs (e.g. the kidneys) are
also in poor shape that the partial defect 
of the liver eventually
proves fatal. It is in this sense that death occurs as
as a kind of overall collapse.
\qpar

The model predicts
that the slope of the death rate is higher
(whether for fall or increase) when the 
number of organs is larger. Can this conjecture
be tested on biological systems? 
The answer is ``yes'' and ``no''. \qL
``Yes'', because a comparison of various species shows that
there are indeed great differences in the number of 
organs. For instance, rotifers (a small swimming animal about
200 micrometers in length) 
have no heart, no blood, no lungs, no kidneys. In short,
they have much less organs than fish.\qL
``No'' because there is a serious obstacle, namely
the great
difference in lifetimes.
Whereas rotifers live about 5 days,  zebrafish live
about 5 years. Unless one knows how to normalize
the respective times there can be no meaningfull comparison
for, needless to say, age normalization 
affects the measurement of the slope. \qL

\appendix

\qI{Appendix A. Statistical versus probabilistic descriptions
of the death process}

In this appendix we discuss some particular aspects of the
theoretical framework. The first subsection clarifies the
connection between age and time.in aging processes.
The second subsection 
establishes the important connection between the density function 
of the ages of death and its frequency counterpart commonly refered to
as the death rate.

\qA{Age versus time}

Should we use age (noted $ x $)
or time (noted $ t $)? At first sight, the question may
seem irrelevant for if time is counted from the
moment of birth the two variables are identical.
It is
not so simple however, as shown by the fact that
in our min-max
argument we had to introduce as many age of death
variables $ X_1, X_2,\ldots $
as there are vital organs. Moreover,
for any organ, its real age is the time elapsed since
it was created. For instance, in zebrafish the heart
appears and starts to beat some 20 hours after fertilization.
Most vital organs are created during embryogenesis
which means that their age 
is not identical with time (measured after
birth which for fish  means hatching of the eggs).
That is why $ X $ was a more appropriate.variable 
than time.
\qpar

However,
in this appendix our perspective is different
for we wish to consider the evolution of
a whole population or more precisely of
a cohort of individuals born at the same moment. Taking
this moment
as origin of the time axis makes the age
of each individual numerically identical with the time
given by an external clock.
\qpar

Actually, to describe the evolution of a cohort, time seems
a better variable than its age. Why? \qL
Although the population exists at any time, for external
observers, it becomes real only when we can know its
size and that occurs only when a census (or a survey) takes
place. Censuses are conducted at specific time intervals 
(e.g. in the US every decade) and concern simultaneously
all cohorts. That is why in this part
calendar time seems to be the natural variable. 
Age will play a role
only if we wish to consider different age groups.

\qA{Death rate versus probability density}

As always, the tricky point is the relationship between
the probabilistic notions and their statistical counterpart. 
The goal of this appendix is to recall the 
main notions and how they are related.
\qpar

Let $ t $ denote the age of individuals in a cohort\qL
Let $ y(t) $ denote the size of the cohort at time $ t $
We wish to describe the decrease of the population
in the course of time.
\qpar

The probability that an individual
would die in the time interval $ (t_1,t_2) $ (which is also 
an age interval) is:
$$ \hbox{(Number of those who die)}/\hbox{(number of those alive initially)}
= [y(t_1)-y(t_2)]/y(t_1) $$

to get the probability of dying per unit of time
we must divide by the length of the time interval 
$ \Delta t=t_2-t_1 $.

 $$ \mu(t)=(1/\Delta t)\
\left[\left( y(t_1)-y(t_2)\right)/y(t)\right] $$

Note that
$ \mu(t) $ represents what is usually called the death rate,
sometimes also called the hazard rate or the force of death.
Note also that: $ y(t_1)-y(t_2) =-\Delta y $.
\qpar

Now, let us consider the case of a constant probability of dying.
One is led to: $ (1/y)(dy/dt)=-a $
which gives: $ y(t)=y_0 \exp(-at) $.
In other words, the survival function is a decreasing exponential.
\qpar

Now let us consider a random variable $ T $ which represents
the age of death of an individual.
Its density function, defined by: $ f(t)dt=P\{t<T<t+dt\} $,
is the derivative of the distribution function: 
$ F(t)=P\{ T\le t \} $.
$ f(t)dt $ is the probability that the death of the
individual occurs
in the age interval $ (t,t+dt) $ ; $ f(t) $ is the 
probability per unit of time. 
\qpar

Note that: $ f(t) \sim \mu(t) $. In words, $ \mu(t) $
is the statistical counterpart of
the probability density function of $ T $.
\qpar

If we consider again the case of a constant probability of dying
(for ages in a bounded interval and zero elsewhere), 
it means: $ f(t)=\mu(t)=a $.
Then, the distribution function is: $ F(t)=at $, at least 
until $ at $ is equal to 1.
The decreasing distribution function is:
 $$ G(t)=P\{ T>t \}=1-F(t)=1-at $$ 

Note that $ G(t) $ is different from the survival function.

\vskip 3mm

{\bf Ethical statement}
\qee{1} The authors did not receive any funding.
\qee{2} The authors do not have any conflict of interest.
\qee{3} The study is purely theoretical and does not involve any
experiment with animals that would require ethical approval.
\qee{4} The study does not involve any participants that
would have to give their informed consent.

\vskip 7mm

{\bf References}

\qpar
1. Berrut,S., Pouillard,V., Richmond,P.,Roehner,B.M.: 
Deciphering infant mortality.
Physica A 463, 400-426 (2016).

\qpar
2. Bois,A., Garcia-Roger,E.M., Hong,E., Hutzler,S.,
Irannezhad,A., Mannioui,A., Richmond,P.,
Roehner,B.M., Tronche,S.:
Infant mortality across species.
A global probe of congenital abnormalities.
Physica A 535, 122308,1-33 (2019).

\qpar
3. Gompertz,B.: On the nature of the function
expressive of the law of human mortality, and on the
mode of determining the value of life contengencies.
Philosophical Transactions of the Royal Society 115,5,513-585
(1825).

\qpar
4. Gavrilov,L.A., Gavrilova,N.S.: The reliability theory
of aging and longevity. 
Journal of Theoretical Biology 213,527-545 (2001).

\qpar
5. Gavrilov,L.A., Gavrilova,N.S.: Reliability theory
of aging and longevity. Chapter 1 (p.3-42) of:
Masoro,E.J., Austad,S.N. editors: Handbook in the 
biology of aging. Elsevier Academic Press, San Diego (CA) 
(2006).

\qpar
6. Baione,F., Levantesi,S.: Pricing critical illness insurance from
prevalence rates: Gompertz versus Weibull. North American Actuarial
Journal, 22,2,270-288 (2018).

\qpar
7. Belayet,H.:  Maternal empowerment and child malnutrition in
Bangladesh. 
Applied Economics 52,14,1566-1581 (2020).

\qpar
8. Elmahdy,E.E.:  A new approach for Weibull modeling for
reliability life data analysis. 
Applied Mathematics and Computation 250,708-720.

\qpar
9. Ricklefs,R.E., Scheuerlein,A.: Biological implications 
of the Weibull and Gompertz models of aging.
The Journals of Gerontology 57,2,1,B69-B76 (2002).

\qpar
10. Chen,Q., Di,Z., Garcia-Roger,E.M., Li,H., Richmond,P.,
Roehner,B.M.: Magnitude and significance of the peak
of early embryonic mortality.
Journal of Biological Physics 17 August (2020).

\qpar
11. Bois,A., Garcia-Roger,E.M., Hong,E., Hutzler,S.,
Irannezhad,A., Mannioui,A., Richmond,P.,
Roehner,B.M., Tronche,S.:
Physical models of infant mortality: implications for 
defects in biological systems.
Journal of Biological Physics 46,371-394 (2020).

\qpar
12. Richmond,P., Roehner,B.M.:
Predictive implications of Gompertz's law.
Physica A 447,446 (2016).

\end{document}